\documentclass[aps,prd,reprint,preprintnumbers,floats,epsfig,nofootinbib,amssymb,
nofootinbib]{revtex4}

\usepackage{slashed}
\usepackage{comment}
\usepackage{graphicx,color}
\usepackage{epsfig}
\usepackage{subfigure}
\usepackage{epsfig}
\usepackage{epstopdf}
\usepackage{dcolumn}
\usepackage{bm}
\usepackage{color}
\usepackage{ulem}
\usepackage{enumitem}

\DeclareUnicodeCharacter{202F}{\,}
\DeclareUnicodeCharacter{2212}{-}
\DeclareUnicodeCharacter{03BC}{\ensuremath{\mu}}

\usepackage[colorlinks,
linkcolor=black,
anchorcolor=black,
citecolor=black
]{hyperref}

\usepackage{orcidlink}

\begin{document}

\baselineskip=15pt


\title{The $U(1)_{L_\mu - L_\tau}$ model meets the new $(g-2)_\mu$ data\\
	and muon neutrino trident scattering}

\author{$^{1,2}$Ming-Wei Li\,\footnote{limw2021@sjtu.edu.cn}\orcidlink{0000-0003-2522-9795}}
\author{$^{1,2}$Xiao-Gang He\,\footnote{hexg@sjtu.edu.cn}\orcidlink{0000-0001-7059-6311}}
\author{$^{1,2}$Andrew Cheek\,\footnote{acheek@sjtu.edu.cn}\orcidlink{0000-0002-8773-831X}}
\author{$^{1,2}$Xinhui Chu\,\footnote{sjtu3915211@sjtu.edu.cn}\orcidlink{0009-0005-0989-7816}}

\affiliation{${}^{1}$State Key Laboratory of Dark Matter Physics,  Tsung-Dao Lee Institute, Shanghai Jiao Tong University, Shanghai 201210, China}
\affiliation{${}^{2}$Shanghai Key Laboratory for Particle Physics and Cosmology, Key Laboratory for Particle Astrophysics and Cosmology (MOE), School of Physics and Astronomy, Shanghai Jiao Tong University, Shanghai 201210, China}

\begin{abstract}
The Muon $g-2$ collaboration at Fermilab has announced their final result of the anomalous magnetic moment of the muon. By adopting the lattice-QCD evaluation of the leading-order hadronic-vacuum-polarization, this result is now in agreement with the latest theoretical prediction to the  $1\sigma$ level. This new result further constrains the allowed parameter space, but does not rule out all possible new physics contributions the muon $g-2$. We study the implications for one of the relevant models, the gauged $U(1)_{L_\mu - L_\tau}$. When using this model to resolve the previous $4\sigma$ tension, results from muon neutrino trident (MNT) scattering experiments would restrict the mass of the new gauge boson ($Z'$) to be less than $300$ MeV. Since the theory and experimental data difference for muon $g-2$ is lowered down to $1\sigma$, the requirement for $m_{Z'}\lesssim 300\,{\rm MeV}$ is much relaxed. Within the updated allowed range of $Z'$ boson mass, we study the models implications for electron and tauon $g-2$ as well as future muon colliders. We find that muon collider can effectively probe the $U(1)_{L_\mu - L_\tau}$.
\end{abstract}

\maketitle

\noindent

\noindent{\bf Introduction} 

The Muon $g-2$ collaboration at Fermilab recently reported their final result for the muon anomalous magnetic dipole moment~\cite{Muong-2:2025xyk} $a_\mu =1165920705(148)\times 10^{-12}$, consistent with previous measurements~\cite{Muong-2:2006rrc, Muong-2:2021ojo, Muong-2:2023cdq}.  Consequently the new world average for muon $\left(g-2\right)_\mu$ becomes $1165920715(145)\times 10^{-12}$. Using the new theory result of Ref.~\cite{Aliberti:2025beg} $a^{\rm SM}_\mu = 116592033(62)\times 10^{-11}$, the difference between experimental data $a^{\rm exp}_\mu$ and new standard model (SM) prediction is now within 1$\sigma$,  $\Delta a_\mu({\rm new}) = a^{\rm exp}_\mu({\rm new}) - a^{\rm SM}_\mu({\rm new}) = (39\pm 64)\times 10^{-11}$. This difference is driven by adopting the lattice-QCD evaluation of the leading-order hadronic-vacuum-polarization which is now much more precise than the data-driven dispersive evaluations~\cite{Aliberti:2025beg}. Hence, the situation now is very different to the previously claimed 4$\sigma$ deviation where $\Delta a_\mu({\rm old}) = a^{\rm exp}_\mu({\rm old}) - a^{\rm SM}_\mu({\rm old}) = (251\pm 59)\times 10^{-11}$. Many theoretical ideas that tried to address the previously claimed deviation now need reevaluation. In this work we study the implications of the new results both in theory and experimental data for the interesting $U(1)_{L_\mu - L_\tau}$ model~\cite{He:1990pn, He:1991qd, Foot:1994vd}  whose $Z'$ gauge boson coupling to muon has a natural contribution to muon $g-2$~\cite{Baek:2001kca, Ma:2001md, Altmannshofer:2014pba, Altmannshofer:2016oaq,Bauer:2018onh}. In this model the same $Z^\prime$ interaction with muons will modify the muon neutrino trident (MNT) scattering $\nu_\mu N \to \nu_i \mu \bar \mu N$. When trying to solve the 4$\sigma$ discrepancy, MNT experiments restricted the $Z^\prime$ mass to be below about $300$ MeV~\cite{Altmannshofer:2014pba,Bauer:2018onh}. Mechanisms had been proposed to evade the MNT constraint and to open the large $Z^\prime$ mass~\cite{Cheng:2021okr}. Since now the discrepancy has disappeared, larger $m_{Z'}$ values for the $U(1)_{L_\mu - L_\tau}$ model are better motivated. We study the new allowed parameter space for the $U(1)_{L_\mu - L_\tau}$ model with constraints from the MNT and the new muon $g-2$ result. We then assess the potential for measurements of the tauon $g-2$ as well as future muon colliders.
\\

\noindent{\bf Muon $g-2$ and $U(1)_{L_\mu - L_\tau}$ } 

In the $U(1)_{L_\mu - L_\tau}$ model, the left-handed $SU(3)_C\times SU(2)_L\times U(1)_Y$ doublets $L_{L\;i}: (1, 2, -1/2)$ and the right-handed singlets $e_{R\;i}: (1,1,-1)$ transform under the gauged $U(1)_{L_\mu-L_\tau}$ group as $0,\;1,\;-1$ for the first, second and third generations, respectively. The $Z^\prime$ gauge boson of the model can receive a mass $m_{Z'} = \tilde g n v_s$ from the vacuum expectation value $\langle S\rangle = v_s/\sqrt{2}$ of a new scalar field $S$ which is a singlet in the SM gauge group, it has a charge of $n$ under the new $U(1)_{L_\mu - L_\tau}$ gauge field. At tree-level, the $Z'$ only interacts with leptons in the  weak interaction basis~\cite{He:1990pn, He:1991qd, Foot:1994vd}
\begin{eqnarray}
\mathcal{L}_{\rm int}=- \tilde g (\bar \mu \gamma^\mu \mu - \bar \tau  \gamma^\mu \tau + \bar \nu_\mu \gamma^\mu L \nu_\mu - \bar \nu_\tau \gamma^\mu L \nu_\tau) Z^\prime_\mu \;, \label{eq:zprime-current}
\end{eqnarray}
where $L(R) = (1 - (+)\gamma_5)/2$. 

\begin{figure}[!t]
	\begin{center}
	\includegraphics[width=\linewidth]{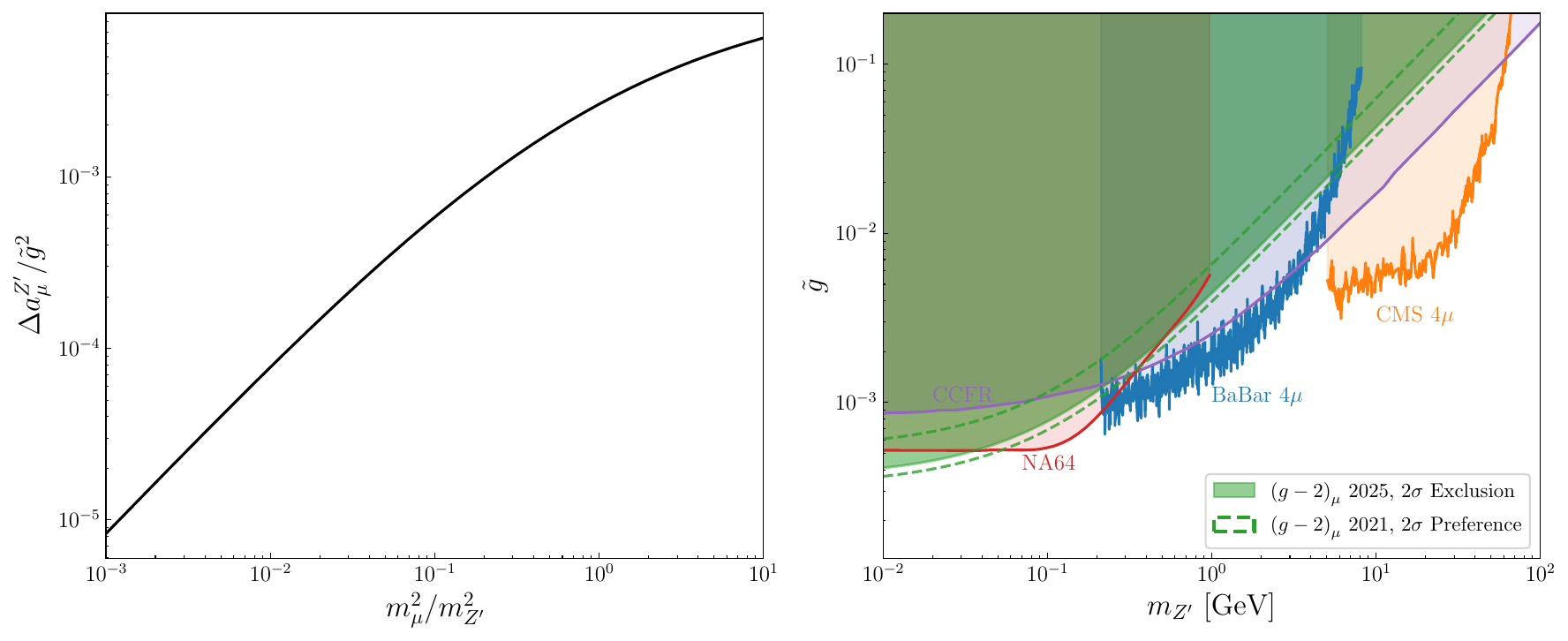}
	\caption{ \textbf{Left panel}: $\Delta a_\mu^{Z^\prime} /\tilde g^2$ as a function of $m^2_\mu/m^2_{Z'}$. \textbf{Right panel}: $\tilde{g}\,\,{\rm vs.}\,\,m_{Z^\prime}$ plane with shaded exclusion regions from Muon $g-2$ 2025 measurement (green)~\cite{Muong-2:2025xyk}, BaBar (blue)~\cite{BaBar:2016sci}, CMS (orange)~\cite{CMS:2018yxg}, NA64 (red)~\cite{NA64:2024klw} and CCFR (purple)~\cite{CCFR:1991lpl}. We also  show the Muon $g-2$ 2021 best fit region between the two dashed green lines.  }
    \label{fig:new_gminus2}
	\end{center}
\end{figure}

Exchanging  $Z^\prime$ at one loop level can generate muon $g-2$~\cite{Baek:2001kca}
\begin{eqnarray}
\Delta a_\mu^{Z^\prime}= {\tilde g^2\over 8\pi^2} {m^2_\mu \over m^2_{Z^\prime}} \int^1_0 {2 x^2(1-x) dx \over 1-x + (m^2_\mu/m^2_{Z^\prime}) x^2}\;. \label{leading}
\end{eqnarray}
In the limit $m_Z'\ll m_\mu$, $\Delta a^{Z^\prime}_\mu$ is given by $\tilde g^2/8\pi^2$ and in the limit $m_{Z^\prime} \gg m_\mu$, 
$\Delta a^{Z^\prime}_\mu = (\tilde g^2/ 12 \pi^2)(m^2_\mu/m^2_{Z^\prime} )$. In the left panel of FIG.~\ref{fig:new_gminus2}, we give $\Delta a^{Z^\prime}_\mu/\tilde g^2$ as a function of $m^2_\mu/m^2_{Z'}$.

To solve the previously claimed 4$\sigma$ deviation of the muon $g-2$ anomaly of 
$\Delta a_\mu({\rm old}) $,  for $m_{Z^\prime} \gg m_\mu$, one would require
${\tilde g^2/m^2_{Z'}}({\rm old}) = (2.66\pm 0.63)\times 10^{-5} \,\mbox{GeV}^{-2}$.
Using the new value for $\Delta a_\mu({\rm new})$ and the muon mass value from Ref.~\cite{ParticleDataGroup:2024cfk}, one would just need ${\tilde g^2/m^2_{Z'}}({\rm new}) = (4.14 \pm 6.79)\times 10^{-6}\, \mbox{GeV}^{-2}$.

For a fixed $\Delta a_\mu$ when the $Z'$ mass becomes larger, the corresponding coupling $\tilde g$ also becomes larger. In order for the theory be perturbative, there is an upper bound for $\tilde g$. For example, requiring $\tilde g^2$ to be less than one, with the new central value of $\Delta a_\mu$, $m_{Z'}$ is limited to be smaller than $490$ GeV.
\\

\noindent{\bf Muon $g-2$ and MNT} 

The $Z^\prime$ interaction of the $U(1)_{L_\mu - L_\tau}$ will also modify the cross section for the MNT scattering 
$\nu_\mu N \to \nu_i \mu \bar \mu N$, compared with the SM prediction by exchanging $Z'$.
For a large $m_{Z'}$ with GeV scale, one can find~\cite{Altmannshofer:2014pba}
\begin{eqnarray}
R=	\left.{\sigma_{Z^\prime}\over \sigma_{\mathrm{SM}}} \right\vert_{\rm trident}= { (1+4s_W^2 + 8 \tilde g^2 m^2_W/g^2 m^2_{Z^\prime})^2 + 1 \over 1 + (1+4 s^2_W)^2}\;, \label{trident-neutrino}
\end{eqnarray}
where $g$ is the $SU(2)_L$ coupling constant and $s_W = \sin\theta_W$ with $\theta_W$ being the Weinberg weak mixing angle.
For small $Z'$ mass of order less than GeV, the behavior does not follow the simple $\tilde g^2/m^2_{Z'}$ scaling and a more careful treatment would be required.
However, some qualitative properties seen in eq.~(\ref{trident-neutrino}) are true for the $m_{Z^\prime}$ values. For example, the $U(1)_{L_\mu-L_{\tau}}$ model only provides constructive interference, see Refs.~\cite{Brown:1972vne,Ballett:2019xoj} for details.
For our purpose we consider large $Z'$ mass above $\sim 60$ GeV, then the formula above is sufficient. Furthermore for such masses, the existing constraints from colliders on this model such as CMS~\cite{CMS:2018yxg} and BaBar~\cite{BaBar:2016sci} lose sensitivity.

Several experiments give the measurements for the ratio $R=\sigma_{ {\rm exp}}/\sigma_{\mathrm{SM}}|_{\rm trident}$: $1.58\pm 0.64$, $0.82\pm 0.28$ and $0.72^{+1.73}_{-0.72}$ from CHARM-II~\cite{CHARM-II:1990dvf}, CCFR~\cite{CCFR:1991lpl} and NuTeV~\cite{NuTeV:1999wlw}, respectively. Notice that for each experiment, the uncertainty allows for $R<1$, in fact the central value for values for both CCFR and NuTeV are below 1. If the results of future experiments such as neutrino near detectors~\cite{Ballett:2018uuc} and forward physics experiments~\cite{Altmannshofer:2024hqd} retrieve a precise determination that $R<1$, the $U(1)_{L_\mu-L_{\tau}}$ will not be able to explain it. Models that yield destructive interference in MNT could explain such a scenario~\cite{Magill:2017mps,Cheng:2022jyi}. 

In the right panel of FIG.~\ref{fig:new_gminus2} we show the most constraining MNT experiment, CCFR, along with the CMS~\cite{CMS:2018yxg}, BaBar~\cite{BaBar:2016sci}, and NA64 (red)~\cite{NA64:2024klw} constraints. In addition we show the Muon $g-2$ 2021 best-fit region by two green dashed lines. Under the previous interpretation of a tension with the SM, this measurement required a significant contribution from the $Z^\prime$ loop and so parameter values between the two lines were the focus of many studies. One can see that at around $m_{Z^\prime}\approx300\,{\rm MeV}$ the CCFR constraint covers the best-fit region. For Babar this occurs around $200\,{\rm MeV}$. Interestingly last year the NA64 result~\cite{NA64:2024klw} large portions of this parameter space, but not all of it. The exclusion region that one gets when taking the 2025 result from the Muon $g-2$ experiment and the new theory prediction for $a_\mu$ is shown with a green shaded region. 

Since now the result is a constraint and not a best-fit region, we are less restricted when choosing a $m_{Z^\prime}$ value. Now there is no preferred parameter space for this model and we should explore all that is allowed. There is a low $m_{Z^\prime}$ limit coming from $N_{\rm eff}$~\cite{Escudero:2019gzq,Holst:2021lzm} and white-dwarf cooling constraints~\cite{Bauer:2018onh,Foldenauer:2024cdp}, requiring that $m_{Z^\prime}\gtrsim 5\,{\rm MeV}$ and there is no theoretical restriction on how heavy $Z^\prime $ can be.

One can see in FIG.~\ref{fig:new_gminus2} that for $m_{Z^\prime}\gtrsim60\,{\rm GeV}$ CCFR provides the strongest constraint, despite not being very precise. Taking a weighted average of CHARM-II, CCFR and NuTeV, we obtain the central value and 1$\sigma$ error as ${\sigma_{ {\rm exp}}/\sigma_{\mathrm{SM}}|_{\rm trident} =0.93\pm 0.25}$. Using the values ${\tilde g^2/m^2_{Z'}}({\rm old}) = (2.66\pm 0.63)\times 10^{-5}\, \mbox{GeV}^{-2}$ and ${\tilde g^2/m^2_{Z'}}({\rm new}) = (4.14 \pm 6.79) \times 10^{-6}\,\mbox{GeV}^{-2}$, one would obtain the following ranges, respectively
\begin{eqnarray}
&&R({\rm old})=\left.{\sigma_{Z^\prime}({\rm old})\over \sigma_{\mathrm{SM}}} \right\vert_{\rm trident} =5.93\pm 1.71\;,\nonumber\\
&&R({\rm new}) = \left.{\sigma_{Z^\prime}({\rm new})\over \sigma_{\mathrm{SM}}} \right\vert_{\rm trident} = 1.47\pm 0.86\;.
\end{eqnarray}

We see that the central value of $R({\rm old})$ is larger than the new value by more than 2$\sigma$. Furthermore, the new value is consistent (within 1$\sigma$ error) with our MNT weighted average.

\vspace{1em}

\noindent{\bf Sub-GeV $Z^\prime$}

As mentioned above, prior to the new theoretical result for the anomalous magnetic moment of the muon, there was much effort to probe the sub-GeV window for the $U(1)_{L_\mu-L_\tau}$ model. The recent result from NA64 went some way to doing this as shown in FIG.~\ref{fig:new_gminus2}. In addition there are many more proposals or experiments that will continue to explore this region of parameter space. This includes using astronomical observations of white dwarfs~\cite{Foldenauer:2024cdp}, future muon beam fixed target experiments such as M$^3$~\cite{Kahn:2018cqs}, and measurements of coherent elastic neutrino-nucleus scattering, both at spallation source experiments~\cite{Miranda:2020syh,Banerjee:2021laz} and in direct dark matter experiments~\cite{Amaral:2021rzw}.

\vspace{1em}

\noindent{\bf The tauon and electron $g-2$}

 In this model $Z'$ exchange will also contribute to the tauon $g-2$. For large $Z'$ mass, we have $\Delta a_\tau = (m^2_\tau / m^2_\mu) \Delta a_\mu = (7.4\pm 18.8)\times 10^{-7} $.  The accuracy of experimental measurements of the tau $g-2$ is comparatively poor with the 95\% allowed range being  $-0.0022<a_\tau<0.0041$~\cite{CMS:2024qjo}. The $Z'$ contribution allowed by the $(g-2)_\mu$ would subsequently induce a $\Delta a_\tau$ that is well within the bound. On the other hand, if one considers the $U(1)_{L_\mu - L_e}$ model, there is $Z'$ exchange contribution to the electron $g-2$. One would have $\Delta a_e=(m^2_e/m^2_\mu)\Delta a_\mu=(9\pm 15)\times10^{-15}$. The experimental value for $a_e$ is $\left(1159.65218059\pm 0.00000013\right)\times 10^{-6}$~\cite{Fan:2022eto}. Again the modification from $Z'$ is well below the current sensitivity.
\\

\noindent{\bf Muon collider signatures}

An additional $Z'$ boson exchange will modify the cross section for muon collisions. In particular for the $U(1)_{L_\mu-L_{\tau}}$ model $\mu \bar \mu \to \tau \bar \tau$ will be the most consequential due to the resonant tree-level exchange of the $Z^\prime$. The scattering amplitude given by
\begin{eqnarray}
&&M(\mu \bar \mu \to \tau \bar \tau) = \left({e^2\over s} - {\tilde g^2 \over s- m^2_{Z'}+ i \Gamma_{Z'} m_{Z'}} \right) \bar \mu \gamma^\mu \mu \bar \tau \gamma_\mu \tau  \nonumber\\
&&+ {g^2/4 \cos^2\theta_W \over s - m^2_Z + i \Gamma_Z m_Z} \bar \mu \gamma^\mu (g_V-g_A \gamma_5)\mu  \bar \tau \gamma_\mu (g_V - g_A \gamma_5 )\tau\;,
\end{eqnarray}
where $e$ is the electric charge, $g$ is the $SU(2)_L$ gauge group coupling, $\theta_W$ is the weak mixing angle, $g_V = -1/2 + 2 \sin^2\theta _W$, and $g_A = -1/2$.

Neglecting muon and tauon masses, we obtain the cross-section as
\begin{eqnarray}
    \sigma\left(\mu \bar \mu \to \tau \bar \tau\right) &=& \frac{4 \pi \alpha^2}{3 s}
+ \frac{\tilde{g}^2 \left( s \left( \tilde{g}^2 - 8 \pi \alpha \right) + 8 \pi \alpha m_{Z'}^2 \right)}{12 \pi \left( (s - m_{Z'}^2)^2 + m_{Z'}^2 \Gamma_{Z'}^2 \right)}
+\frac{G_F^2 \left( g_A^2 + g_V^2 \right)^2 s\, m_Z^4}{6 \pi \left( (s - m_Z^2)^2 + m_Z^2 \Gamma_Z^2 \right)}\nonumber\\
&& +\frac{2 \sqrt{2}\, \alpha\,G_F\,m_Z^2\, g_V^2\, \, (s - m_Z^2)}{3 \left( (s - m_Z^2)^2 + m_Z^2 \Gamma_Z^2 \right)}\left(1-\frac{\tilde{g}^2\, s \left( (s - m_Z^2)(s - m_{Z'}^2)+ m_Z\, m_{Z'}\, \Gamma_Z\, \Gamma_{Z'} \right)}{4 \pi \alpha (s - m_Z^2)\left( (s - m_{Z'}^2)^2 + m_{Z'}^2\, \Gamma_{Z'}^2 \ \right)}
\right)
\end{eqnarray}
In the above we have removed $g^2$ by the substitution $g^2=4\sqrt{2} \,m_Z \,G_F\cos^2\theta_{\rm W}$. The decay width $\Gamma_{Z'}$ of $Z'$ can be calculated from eq.~\ref{eq:zprime-current} summing over muon and tauon pairs and their corresponding neutrinos to obtain $\Gamma_{Z'} = \tilde g^2 m_{Z'}/4 \pi$.

We now discuss the prospects of detecting deviations from the SM in $\mu \bar \mu \to \tau \bar \tau$ scattering measured at a future muon collider. We take the benchmark choices of Ref.~\cite{Delahaye:2019omf}, in particular we take the scaling of the integrated luminosity of a muon collider between energies $1-3$ TeV to be $(\sqrt{s})^2$. For a 5 year runtime, the integrated luminosity for $\sqrt{s} = 1$ TeV ($\sqrt{s} = 3$ TeV) is projected to be roughly $1\,{\rm ab}^{-1}$ ($0.11 \,\rm{ab}^{-1}$) which corresponds to $1.09\times 10^{5}$ ($1.39\times 10^{3}$) tauon pair events. For comparison with the SM, we use the most recent values from Ref.~\cite{ParticleDataGroup:2024cfk}.

\begin{figure}
    \centering
    \includegraphics[width=\linewidth]{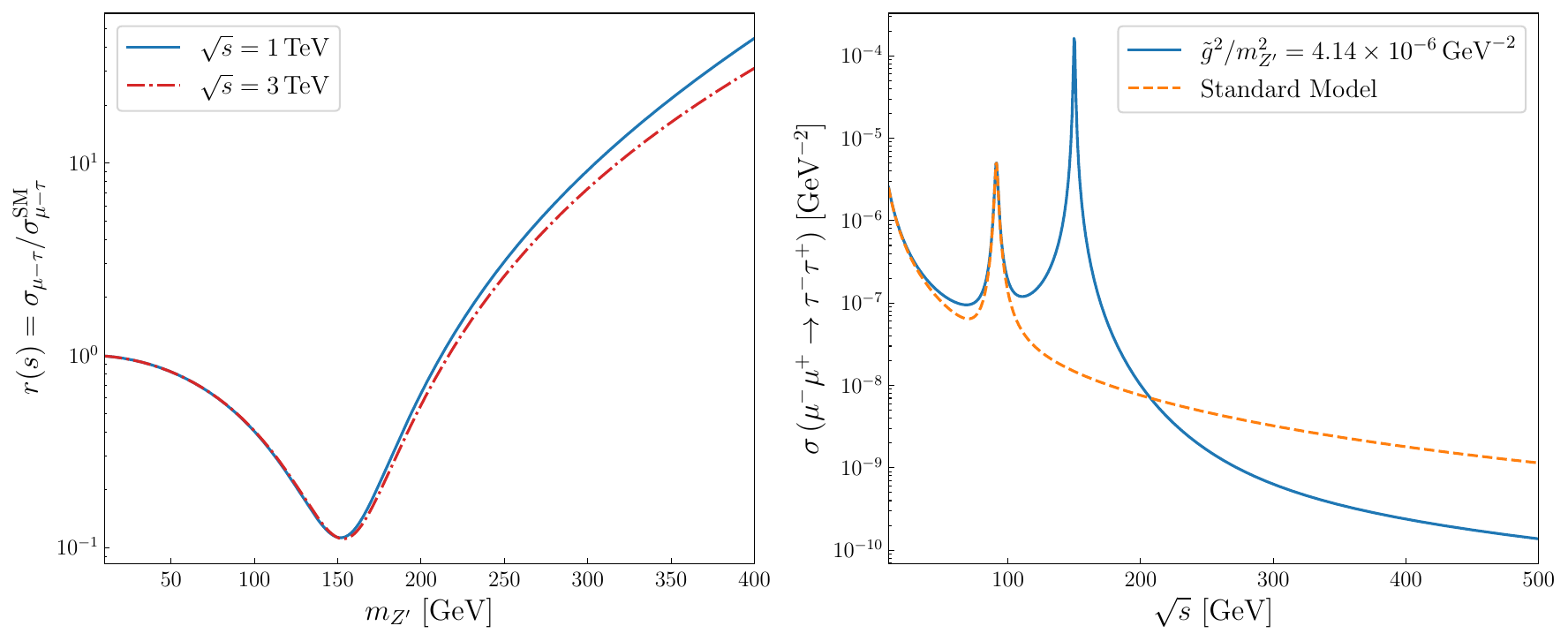}
    \caption{\textbf{Left panel}: The $\mu \bar \mu \to \tau \bar \tau$ process ratio $r=\sigma_{\mu-\tau}/\sigma^{\rm SM}_{\mu-\tau}$ as a function of $m_{Z'}$, from 10 GeV to 400 GeV. We set $\tilde g^2 /m^2_{Z'} = 4.14\times 10^{-6}/\,{\rm GeV}^2$ and show $\sqrt{s}=1\,{\rm TeV}$ (blue solid line) and $3\,{\rm TeV}$ (red dot-dashed line). \textbf{Right panel}: The $\mu \bar \mu \to \tau \bar \tau$ cross section as a function of $\sqrt{s}$ assuming the SM only (dashed orange line) and the $U(1)_{L_\mu - L_\tau}$ model (solid blue line) between 10 GeV and 500 GeV. We also take $\tilde g^2 /m^2_{Z'} = 4.14\times 10^{-6}/\,{\rm GeV}^2$ and $m_{Z'} = 150$ GeV.}
    \label{fig:collider_combined}
\end{figure}

In figure~\ref{fig:collider_combined}, on the left panel, we plot the cross section $\sigma_{\mu-\tau}$ of $\mu \bar \mu \to \tau \bar \tau$ divided by the $\sigma^{\rm SM}_{\mu -\tau}$, $r = \sigma_{\mu-\tau}/\sigma^{\rm SM}_{\mu-\tau}$ for two values of $\sqrt{s}=$ 1 TeV (solid blue line) and 3 TeV (dot-dashed red line) for $m_{Z'}$ varying from 10 GeV to 400 GeV. One sees that for $m_{Z'}$ mass below 200 GeV, the event number decreases as much as factor of 9, but above 200 GeV increases to more than 10 times compared with the SM predictions. Such large differences can be detected even with modest tauon pair reconstruction, say 10\%. This can be used to determine if $Z'$ in the model couples to muon and tauon as predicted in the $U(1)_{L_\mu - L_\tau}$ model. If the decreasing or increasing effects are not seen one can constrain $\tilde g^2/m^2_{Z^\prime}$ to be smaller by several times. On the right panel of figure~\ref{fig:collider_combined}, we also take a $m_{Z'} = 150\,{\rm GeV}$ for example to see how the cross section changes with $\sqrt{s}$ across the resonant point. The effects are also significant. This emphasizes the impact a future muon collider will have on tests of the $U(1)_{L_\mu - L_\tau}$ model. In this paper we focus only on the 2$\to$2 scattering process which is typically dominant and easy to analyze, processes with initial state radiation can also provide important channels when one considers a full collider analysis~\cite{Huang:2021nkl}.
\\

\noindent{\bf Conclusion}

We study the implications of the recent reassessment of the theoretical prediction for the anomalous magnetic moment of the muon and the experimental determination reported by the the Muon $g-2$ collaboration. We do this specifically for one of the simplest extensions to the SM, the $U(1)_{L_\mu - L_\tau}$ model. The previous emphasis on solving the tension between theory and experiment for $(g-2)_\mu$ restricted the $Z'$ mass to be less than 300 MeV. This was because the required parameters would provoke observable consequences in existing MNT experiments.

Now that the SM prediction is consistent with the measurement, there is no such restriction, and one is motivated to consider larger $m_{Z'}$ values. For example $m_{Z'}\gtrsim\mathcal{O}(100\,{\rm GeV})$ remains very much allowed even with the new $\Delta a_\mu$ central value. Furthermore, measurements of the tauon $g-2$ as well as results from existing collider and MNT experiments are fairly weak. This means the gauged $U(1)_{L_\mu - L_\tau}$ is very much alive and well. Therefore, with this new outlook, we show that a future muon collider will be an effective and powerful tool for probing the $U(1)_{L_\mu - L_\tau}$ model.

\section*{Acknowledgments}
This work was partially supported by the Natural Science Foundation of the People’s Republic of China (Nos. 12090064, 12375088, W2441004 and 123B2079).
\bibliographystyle{unsrturl}
\bibliography{refs}  

\end{document}